\documentclass[draftclsnofoot,onecolumn,romanappendices]{IEEEtran}
\usepackage{amsmath,amsfonts,amssymb,amsthm}
\usepackage{url}
\usepackage{cite}
\usepackage{tikz}
\usepackage{graphicx}

\newtheorem{theorem}{Theorem}
\newtheorem{lemma}{Lemma}

\newtheorem{claim}{Claim}
\theoremstyle{definition}
\newtheorem{definition}{Definition}
\newtheorem{example}{Example}
\theoremstyle{mark}

\newcommand{\ein}{\mathcal{E}_{\mathrm{i}}}
\newcommand{\eout}{\mathcal{E}_{\mathrm{o}}}
\newcommand{\tail}{\mathrm{tail}}
\newcommand{\head}{\mathrm{head}}
\newcommand{\mincut}{\textrm{min-cut}}

\newcommand{\bx}{x}

\newcommand{\Fq}{\mathbb{F}_q}

\newcommand{\mA}{\mathcal{A}}
\newcommand{\mE}{\mathcal{E}}
\newcommand{\mN}{\mathcal{N}}
\newcommand{\mC}{\mathcal{C}}
\newcommand{\mO}{\mathcal{O}}
\newcommand{\mV}{\mathcal{V}}

\tikzstyle{vertex}=[draw,circle,fill=gray!30,minimum size=6pt, inner sep=0pt]

\makeatletter

\newcommand{\Rmnum}[1]{\expandafter\@slowromancap\romannumeral #1@}
\makeatother
\begin{document}
\title{Comments on Cut-Set Bounds on Network Function Computation}
\author{Cupjin~Huang, Zihan~Tan, Shenghao~Yang, and Xuan Guang%
  \thanks{This paper was presented in part at 2015 IEEE Information
    Theory Workshop (ITW), Jerusalem, Israel.}
  \thanks{C. Huang is with the Department of Electrical Engineering
    and Computer Science, University of Michigan, Ann
    Arbor, MI, USA. E-mail: cupjinh@umich.edu}
  \thanks{Z. Tan is with the Department of Computer Science, The
    University of Chicago, IL, USA. E-mail: zihantan@uchicago.edu}
  \thanks{S. Yang is with the School of Science of Engineering, The
    Chinese University of Hong Kong, Shenzhen, China. E-mail: shyang@cuhk.edu.cn}
  \thanks{X. Guang is with the School of Mathematical Sciences, Nankai University, Tianjin, and the Institute of Network Coding, The Chinese University of Hong Kong, Hong Kong SAR, China. E-mail: xguang@nankai.edu.cn}

}

\maketitle

\begin{abstract}
  A function computation problem in directed acyclic networks has been considered
  in the literature, where a sink node wants to compute a
  target function with the inputs generated at multiple source
  nodes. The network links are error-free but capacity-limited, and
  the intermediate network nodes perform network coding. The target
  function is required to be computed with zero error. The computing
  rate of a network code is measured by the average number of times
  that the target function can be computed for one use of the
  network, i.e., each link in the network is used at most once. In the papers \cite{Appuswamy11, Kowshik12}, two cut-set bounds
  were proposed on the computing rate. However, we show in this paper
  that these bounds are not valid for general network function computation problems. We analyze the arguments that lead to
  the invalidity of these bounds and fix the issue with a new
  cut-set bound, where a new equivalence relation associated
  with the inputs of the target function is used. Our bound is qualified for general target
  functions and network topologies. We also show that our bound is
  tight for some special cases where the computing capacity is known. Moreover, some results in \cite{Appuswamy13, Appuswamy14} were proved using the invalid upper bound in \cite{Appuswamy11} and hence their correctness needs further justification. We also justify their validity in the paper.
\end{abstract}

\section{Introduction}
We consider a function computation problem in a directed acyclic network, where
a \emph{target function} is intended to be calculated at a sink
node, and the input symbols of the target function are generated at
multiple source nodes. As a special case, network communication is
just the computation of the \emph{identity function} whose output is
the same as the input. Network function computation naturally arises
in sensor networks~\cite{Giridhar05} and has applications in big data processing.
Various models and special cases of this problem have been studied in the
literature (see the summarizations in
\cite{Appuswamy11,Kowshik12,Rai12,Zhang13,Ramam13}).

The following \emph{network coding model} for function computation has
been studied in \cite{Appuswamy11,Kowshik12}. Specifically, the
network links have limited (unit) capacity and are error-free. Each
source node generates multiple input symbols, and the network codes
perform vector network coding by using the network multiple times,
where one use of a network means the use of each link
at most once. An intermediate network node can transmit
a certain fixed function of the symbols it receives. Here all the
intermediate nodes are considered with unbounded computing
capability. The target function is required to be computed correctly for
all possible inputs. We are interested in the \emph{computing rate}
of a network code that computes the target function, i.e., the average
number of times that the target function can be computed for one use
of the network. The maximum computing rate is called the
\emph{computing capacity}.

For a general target function, the computing capacity is known when
the network is a multi-edge tree \cite{Appuswamy11}. For general
network topologies, when the identity function is required to be computed, the problem
becomes the extensively studied network coding \cite{flow,linear}, and
it is known that in general linear network codes are sufficient to
achieve the multicast capacity \cite{linear, alg}. For scalar linear
target functions, the computing capacity can be fully characterized by
min-cut using the duality technique of linear network codes proposed
in \cite{koetter2004network}. For (vector) linear target functions
over a finite field, a complete characterization of the computing
capacity is not available. But certain necessary and sufficient conditions
have been obtained so that linear network codes are sufficient to
calculate a linear target function \cite{Appuswamy14}. %

For the general target functions and network topologies, various upper
bounds on the computing capacity based on cut sets have been studied
\cite{Appuswamy11,Kowshik12}. We find, however, that the upper bounds
claimed in the previous works do not always hold. For an example we will evaluate in Section~\ref{sec:probound}, the computing capacity is strictly larger than
the two upper bounds claimed in \cite{Appuswamy11,Kowshik12}
respectively, where the issue is related to the equivalence relation
associated with the target function defined in these papers.
Characterizations based on this equivalence relation is only valid for
special network topologies, e.g., the multi-edge tree. For more
general networks, however, this equivalence relation is not
sufficient to explore general function computation problems.

Towards a general upper bound, we define a new equivalence relation
associated with the inputs of the target function (but does not depend
on the network topology) and propose a cut-set bound on the computing
capacity using this equivalence relation (see
Section~\ref{sec:general_bound}). The obtained bound holds for general
target functions and general network topologies. We further discuss
how to simplify the bound, and show that this bound is tight for some
special cases where the computing capacity is known.  %

Some results in \cite{Appuswamy13, Appuswamy14} were proved using the
invalid upper bound in \cite{Appuswamy11} and hence their correctness
needs further justification.  The validity of these results in \cite{Appuswamy13,
  Appuswamy14} is justified in Section~\ref{sec:further-comments}.

\section{Issues of the Previous Bounds}
\label{sec:probound}

In this section, we introduce the network computing model and discuss the issues of the previous bounds.

\subsection{Function-Computing Network Codes}
\label{sec:funct-comp-netw}

Let $G=(\mathcal{V},\mathcal{E})$ be a directed acyclic graph (DAG)
with a finite node set $\mathcal{V}$ and an edge set $\mathcal{E}$,
where multi-edges between a pair of nodes are allowed. A {\em
  network} over $G$ is denoted as $\mathcal{N}=(G,S,\rho)$, where
$S\subset \mathcal{V}$ is the set of {\em source nodes}, say $S=\{\sigma_1, \sigma_2, \cdots, \sigma_s \}$ with $|S|=s$, and $\rho\in
\mathcal{V}\backslash S$ is the single {\em sink node}.
For an edge $e=(u,v)$, we call $u$ the tail
of $e$ (denoted by $\tail(e)$) and $v$ the head of $e$ (denoted by
$\head(e)$). Moreover, for each node $u\in\mathcal{V}$, let $\ein(u) =
\{e\in \mathcal{E}: \head(e)=u\}$ and $\eout(u) =
\{e\in\mathcal{E}:\tail(e)=u\}$ be the set of incoming edges and the
set of outgoing edges of $u$, respectively. Fix a topological order ``$>$'' of
all the nodes in $\mathcal{V}$. This order naturally induces an order of
all the edges in $\mathcal{E}$, also denoted by ``$>$'', where for two edges $e$ and $e'$, $e>e'$ if either
$\tail(e)>\tail(e')$ or $\tail(e)=\tail(e')$ and
$\head(e)>\head(e')$. Without loss of generality, we assume that $\ein(\sigma)=\emptyset$ for all
source nodes $\sigma\in S$, and $\eout(\rho)=\emptyset$.

The network defined above is used to compute a function, where
multiple inputs are generated at the source nodes and the output of
the function is demanded by the sink node. The computation units with
unbounded computing capability are allocated at all the network
nodes. However, the computing capability of the network is still
bounded by the network transmission capability. Let $\mathcal{B}$ be a finite alphabet and we assume that each edge can transmit a symbol in $\mathcal{B}$ reliably for each use.

Let $\mathcal{A}$ and $\mathcal{O}$ be two finite alphabets. Let $f:\mathcal{A}^s\to \mathcal{O}$ be the {\em target function}, which is required to be computed via the network and whose $i$th input is generated at the $i$th source node $\sigma_i$. It is allowed to use
the network multiple times to compute the function. Suppose the source node $\sigma_i$ generates $k$ symbols in $\mathcal{A}$, denoted by $x_i=(x_{i,1},x_{i,2},\cdots,x_{i,k})^\top$.
The symbols generated by all
the source nodes can be given as a \emph{message matrix} $x_S=(x_{1}, x_{2}, \cdots, x_{s})$ of size
$k\times s$. %
Let
\begin{equation*}
  f(x_S) = \big(f(x_{1,j},x_{2,j},\cdots,x_{s,j}):\ j=1,2,\ldots,k\big)^{\top}
\end{equation*}
be $k$ outputs of the target function $f$.
For any source subset $J\subseteq S$, we let
$x_J=(x_{i}: \sigma_i\in J)$ and use $\mA^{k\times J}$ (instead of $\mA^{k\times |J|}$) to
denote the set of all possible $k\times |J|$ matrices taken by $x_J$.
In particular, when $J=\emptyset$, we adopt the convention that $x_J$ is empty.
We equate $\mA^{1\times J}$ with $\mA^{J}$.

For two positive integers $k$ and $n$, a $(k,n)$ (function-computing)
network code over a network $\mathcal{N}$ with a target function $f$ is
defined as follows. Let $x_S\in \mathcal{A}^{k\times S}$ be the message matrix formed by symbols generated at the source nodes. The purpose of such a network code is to obtain
$f(x_S)$ at the sink node $\rho$ by transmitting at most $n$ symbols in $\mathcal{B}$ on
each edge in $\mathcal{E}$, say, using the network at most $n$ times.
The $(k,n)$ network code contains a local encoding function for each
edge $e$:
\begin{equation*}
  h_{e}: \left\{
  \begin{array}{ll}
    \mathcal{A}^k \rightarrow \mathcal{B}^n, & \mathrm{if }\ \mathrm{tail}(e)\in S; \\
    \prod_{e'\in \ein(\mathrm{tail}(e))} \mathcal{B}^n \rightarrow
    \mathcal{B}^n, & \mathrm{otherwise.}
  \end{array}
  \right.
\end{equation*}
For each edge $e$, $h_e$ is executed on the tail of $e$ and determines the symbols transmitted on $e$. The execution of these functions $h_e$ follows the order on edges.

Denote the symbols transmitted on an edge $e$
by $g_{e}(x_S) \in \mathcal{B}^n$. For a set of edges
$E\subseteq \mathcal{E}$, we let
\begin{equation*}
  g_{E}(x_S)=\big( g_{e}(x_S): e\in E \big).
\end{equation*}
Similar to the classic network coding \cite{flow,linear}, if $e$ is an outgoing edge of the $i$th source node $\sigma_i$, then
\begin{equation*}
  g_{e}(x_S) = h_{e}(x_i);
\end{equation*}
if $e$ is an outgoing edge of $u\in \mathcal{V}\setminus S$, then
\begin{equation*}
  g_{e}(x_S) = h_{e}\left(g_{\ein(u)}(x_S)\right).
\end{equation*}

The $(k,n)$ network code also contains a decoding function
\begin{equation*}
  \varphi: \prod_{e'\in \ein(\rho)} \mathcal{B}^n \rightarrow
    \mathcal{O}^k
\end{equation*}
at the sink node $\rho$.
Define
\begin{equation*}
  \psi(x_S) = \varphi\left(g_{\ein(\rho)}(x_S)\right).
\end{equation*}
If the network code {\em computes} $f$, i.e., $\psi(x_S)=f(x_S)$ for all message matrices $x_S\in\mA^{k\times S}$, then we call
$\frac{k}{n}\log_{|\mathcal{B}|}|\mathcal{A}|$ an {\em achievable}
computing rate.\footnote{We multiply $\frac{k}{n}$ by
$\log_{|\mathcal{B}|}|\mathcal{A}|$ in order to normalize the computing rate for target functions with
different input alphabets.}
The {\em computing capacity} of the network $\mathcal{N}$ with respect
to the target function $f$ is defined as
$$\mathcal{C}(\mathcal{N},f)=\sup\left\{
  \frac{k}{n} \log_{|\mathcal{B}|}|\mathcal{A}|:\ \frac{k}{n}\log_{|\mathcal{B}|}|\mathcal{A}|\textrm{ is achievable} \right\}.$$

\subsection{Cut Sets}

The upper bounds on the computing capacity discussed in this paper are related to the cut sets of the network~$\mathcal{N}$. So we briefly discuss cut sets and define some notations.

For two nodes $u$ and $v$ in $\mathcal{V}$, if there exists a directed path from $u$ to $v$ in
$G$, we say $v$ is \emph{reachable} by $u$ and denote the relation by $u\rightarrow v$. We adopt the convention that a node is reachable by itself. If there is no directed path from $u$ to $v$, we say that $u$ is
\emph{separated} from $v$. We assume that i) $u\rightarrow \rho$ for all
$u\in\mathcal{V}$ and ii) any node $u \in
\mathcal{V}$ is reachable by at least one source
node.\footnote{We can delete the nodes not satisfying i) or ii)
  from the original network without reducing the computing capacity.}

Given a set of edges $C\subseteq
\mathcal{E}$, $I_C$ is defined to be the set of source nodes which are
separated from the sink node $\rho$ if $C$ is deleted from $\mathcal{E}$, i.e.,
\begin{equation*}
I_C=\left\{ \sigma \in S:\ \sigma \text{ is separated from } \rho \text{ upon deleting the edges in $C$ from $\mathcal{E}$} \right\}.
\end{equation*}
An edge set $C$ is called a {\em
  cut set} if $I_C\neq \emptyset$. The family of all cut sets in the network
$\mathcal{N}$ is denoted by $\Lambda(\mathcal{N})$, i.e.,
$$\Lambda(\mathcal{N})=\{ C\subseteq \mathcal{E}:\ I_C \neq \emptyset \}.$$

Additionally, we define the set $K_C$ as
$$K_C=\left\{ \sigma\in S:\ \exists e\in C \text{ s.t. } \sigma\rightarrow \tail(e) \right\}.$$
We see that $K_C$ is the set of
source nodes from each of which there exists a path to $\rho$ through
$C$, and hence $I_C\subseteq K_C$. Further, let $$J_C=K_C\backslash I_C.$$

We say a subset of nodes $U\subseteq
\mathcal{V}$ is a \emph{cut} if $|U\cap S|>0$ and $\rho \notin U$. For
a cut $U$, denote by $\mathcal{E}(U)$ the cut set induced by $U$,
i.e.,
\begin{equation*}
  \mathcal{E}(U) = \{e\in \mathcal{E}: \tail(e)\in U \text{  and  } \head(e)\in
  \mathcal{V}\setminus U\}.
\end{equation*}
Let $$\bar{\Lambda}(\mathcal{N}) = \{\mathcal{E}(U): U \text{ is a cut in }
\mathcal{N}\}.$$
It is clear that $\bar{\Lambda}(\mathcal{N}) \subseteq \Lambda(\mathcal{N})$.

\begin{lemma}\label{lemma:cc}
  For any non-empty subset of the source nodes $S'\subseteq S$,
  \begin{equation*}
    \min_{C\in \Lambda(\mathcal{N}):I_C=S'} |C| = \min_{C\in \bar{\Lambda}(\mathcal{N}):I_C=S'} |C|.
  \end{equation*}
\end{lemma}
\begin{IEEEproof}
  The lemma is a direct consequence of the max-flow min-cut theorem.
\end{IEEEproof}

\subsection{Invalidity of the Previous Upper Bounds}
\label{sec:inval-prev-upper}

The upper bounds on the computing capacity in \cite{Appuswamy11,
  Kowshik12} are based on the following equivalence relation.
For $x_S \in \mA^{k\times S}$ and a partition $\{I,J\}$ of $S$, we abuse the notations and write $x_S = (x_I,x_J)$ and $f(x_S) = f(x_I,x_J)$.

\begin{definition}\label{def:I-equi}
Consider a function $f:\mathcal{A}^s\to \mathcal{O}$ and a subset $I
\subseteq S$. For any $a_I, b_I \in \mA^I$, we say $a_I$ and $b_I$ are
$I$-equivalent (with respect to $f$) if $f(a_I,d_{S\setminus I}) = f(b_I,d_{S\setminus I})$ for every $d_{S\setminus I} \in \mA^{S\setminus I}$.
\end{definition}

For a target function $f$ and a subset $I\subseteq S$, denote by $R_{I,f}$ the
total number of $I$-equivalence classes. %
Define
\begin{align*}
\mincut_{\text{A}}(\mathcal{N},f)\triangleq & \min_{C\in\Lambda(\mathcal{N})}\dfrac{|C|}{\log_{|\mathcal{A}|}R_{I_C,f}},\\
\mincut_{\text{K}}(\mathcal{N},f)\triangleq & \min_{C\in\bar{\Lambda}(\mathcal{N})}\dfrac{|C|}{\log_{|\mathcal{A}|}R_{I_C,f}}.
\end{align*}
It is claimed in \cite[Theorem~II.1]{Appuswamy11} that
$\mincut_{\text{A}}(\mathcal{N},f)$ is an upper bound on
$C(\mathcal{N},f)$; and it is implied by \cite[Lemma~3]{Kowshik12} that
$\mincut_{\text{K}}(\mathcal{N},f)$ is an upper bound on
$\mathcal{C}(\mathcal{N},f)$. Further, we can show that $\mincut_{\text{K}}(\mathcal{N},f) =
\mincut_{\text{A}}(\mathcal{N},f)$ by writing
\begin{IEEEeqnarray*}{rCl}
  \mincut_{\text{A}}(\mathcal{N},f)=
  \min_{S'\subseteq S} \frac{\min\limits_{C\in \Lambda(\mathcal{N}):I_C=S'} |C|}{\log_{|\mathcal{A}|}R_{S',f}}
=\min_{S'\subseteq S} \frac{\min\limits_{C\in \bar{\Lambda}(\mathcal{N}):I_C=S'} |C|}{\log_{|\mathcal{A}|}R_{S',f}}
=\mincut_{\text{K}}(\mathcal{N},f),
\end{IEEEeqnarray*}
where the second equality follows from Lemma~\ref{lemma:cc}.

Though $\mincut_{\text{A}}(\mathcal{N},f)$ is an upper bound for special cases (e.g., for a tree network with a general target function or a general network with
the identity target function), it is not a valid upper bound in
general. We first use an example to illustrate this fact.

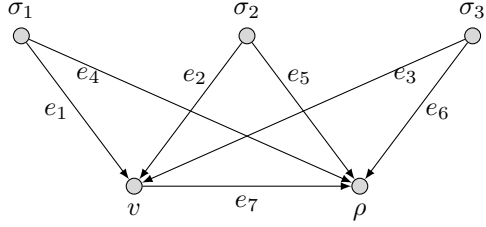
\begin{figure}
  \centering
 \begin{tikzpicture}
    \draw (-3, 0) node[vertex] (1) [label=above:$ \sigma_1$] {};
    \draw (0, 0) node[vertex] (2) [label=above:$ \sigma_2$] {};
    \draw (3, 0) node[vertex] (3) [label=above:$ \sigma_3$] {};
    \draw (-1.5, -2) node[vertex] (v) [label=below:$v$] {};
    \draw (1.5,-2) node[vertex] (r) [label=below:$\rho$] {};
    \draw[->,>=latex] (1) -- (v) node [midway, left] {$e_1$};
    \draw[->,>=latex] (2) -- (v) node [near start, left] {$e_2$};
    \draw[->,>=latex] (3) -- (v) node [near start, right] {$e_3$};
    \draw[->,>=latex] (1) -- (r) node [near start, left] {$e_4$};
    \draw[->,>=latex] (2) -- (r) node [near start, right] {$e_5$};
    \draw[->,>=latex] (3) -- (r) node [midway, right] {$e_6$};
    \draw[->,>=latex] (v) -- (r) node [midway, below] {$e_7$};
\end{tikzpicture}
\caption{Network $\mathcal{N}_1$ has three source nodes, $\sigma_1$, $\sigma_2$ and $\sigma_3$, and one sink node $\rho$ that computes the nonlinear function
  $f(x_1,x_2,x_3)=x_1x_2+x_3$, where $\mathcal{A}=\mathcal{B} = \mathcal{O}=\{0,1\}$.}
  \label{fig:1}
\end{figure}

\begin{example}
Consider the network $\mathcal{N}_1$ in Fig.~\ref{fig:1} with the target function
$f(x_1,x_2,x_3)=x_1x_2+x_3$, where
$\mathcal{A}=\mathcal{B} = \mathcal{O}=\{0,1\}$, regarded as the finite filed $\mathbb{F}_2$.
There exists a $(2,1)$ network code that
computes $f$ in $\mathcal{N}_1$, where the source node $\sigma_i$ sends $x_{i,1}$ to the intermediate node $v$ and sends
$x_{i,2}$ to the sink node $\rho$ for $i=1,2,3$ respectively, i.e., for
$i=1,2,3$,
\begin{equation*}
  g_{e_i}(x_S) =  x_{i,1}, \quad g_{e_{i+3}}(x_S)  =  x_{i,2}.
\end{equation*}
The node $v$ computes $f(x_{1,1},x_{2,1},x_{3,1})$ and sends it to $\rho$ via the edge $e_7$. In addition to receiving $f(x_{1,1},x_{2,1},x_{3,1})$ from $e_7$, $\rho$ computes
$f(x_{1,2},x_{2,2},x_{3,2})$ by using the symbols received from the edges
$e_4, e_5$ and $e_6$. Therefore, $\mathcal{C}(\mathcal{N}_1,f) \geq 2$.

On the other hand, for the cut set
$C_1=\{e_4,e_6,e_7\}$, we have $I_{C_1}=\{ \sigma_1, \sigma_3\}$ and $R_{I_{C_1},f}=4$
because any two out of all possible inputs $(0,0)$, $(0,1)$, $(1,0)$ and $(1,1)$ taken by $(x_1, x_3)$ are not $I_{C_1}$-equivalent.
Hence,
$$\mincut_{\text{A}}(\mathcal{N}_1,f) \leq
\dfrac{|C_1|}{\log_{|\mathcal{A}|}R_{I_{C_1},f}}=\dfrac{3}{2} < 2 \leq \mathcal{C}(\mathcal{N}_1,f),$$
which shows the invalidity of these two bounds.
\end{example}

Next, we explain why the bounds are invalid. In both papers, the following claim is considered as a necessary condition such that a $(k,n)$ network code can compute the target function (see \cite[proof of Theorem II.1]{Appuswamy11} and \cite[Lemma 3]{Kowshik12}).
\begin{claim}
  For any cut set $C$ and $a_{I_C}=(a_{I_C,1}, \ldots, a_{I_C,k})^\top, b_{I_C}=(b_{I_C,1},\ldots,b_{I_C,k})^\top \in \mA^{k\times I_C}$, if for some $j$, $1 \leq j \leq k$,  $a_{I_C,j}$ and $b_{I_C,j}$ are not $I_C$-equivalent (i.e, there exists a $d_{S\setminus I_C} \in \mA^{S\setminus I_C}$ such that $f(a_{I_C,j},d_{S\setminus I_C})\neq f(b_{I_C,j},d_{S\setminus I_C})$), then cut set $C$ should transmit distinct messages with respect to $a_{I_C}$ and $b_{I_C}$.
\end{claim}

In fact, $I_C$-equivalence is not necessary in general for a network code to
compute the target function. The point is that when encoding the
transmitted messages for the cut set $C$, the messages generated by
the source nodes in $S\setminus I_C$ are useful as well, so that it is
not necessary to explore all the possible inputs at the source nodes
in $S\setminus I_C$ to find the worst case, i.e., the above claim is
not necessary for a $(k,n)$ network code to compute the target function. %

The $I_C$-equivalence is necessary for some special cases. For instances, in a tree network, the encoded messages transmitted on a cut set $C$ do not depend on the inputs generated by the source nodes in $S\setminus I_C$ since any source node in $S\setminus I_C$ is separated from any node $\tail(e)$, $e\in C$; and for the identity target function, knowing the inputs generated by the source nodes in $S\setminus I_C$ is useless since $f(a_{I_C},d_{S\setminus I_C})\neq f(b_{I_C},d_{S\setminus I_C})$ for any $d_{S\setminus I_C}\in \mA^{S\setminus I_C}$ as long as $a_{I_C}\neq b_{I_C}$.

\section{General Cut-Set Bound}
\label{sec:general_bound}

In this section, we put forward a cut-set bound that is applicable for general network topologies and general target functions.

\subsection{Equivalence Relation}

We first propose a new equivalence relation that can fix the issue of the previous one defined in Definition~\ref{def:I-equi}.

\begin{definition}[Equivalence Relation]\label{def:ec}
  Let $f:\mathcal{A}^s\to \mathcal{O}$ be a target function. Consider two disjoint sets $I,J\subseteq S$ and a fixed $c_J\in\mA^{J}$. For any $a_I,b_I\in \mathcal{A}^{I}$, we say $a_I$ and $b_I$ are
  $(I,J,c_J)$-equivalent (with respect to $f$) if
  $f(a_I,c_J,d) = f(b_I,c_J,d)$ for all $d \in \mA^{S\setminus (I\cup J)}$.
\end{definition}

We can easily see that this relation is an equivalence one. This equivalence relation does not depend on the network topology, and instead, only depend on the function. It will soon be clear that with a network, the division of equivalence classes naturally leads to an upper bound of the computing capacity based on cut sets.

We give several examples of the equivalence relation, where
$I,J\subseteq S$ are disjoint and let
$a_I, b_I\in \mathcal{A}^{I}$ and $c_J\in\mathcal{A}^{J}$.
\begin{example}\label{ex:1}
  For the identity function, $a_I$ and $b_I$ are $(I,J,c_J)$-equivalent if $a_I = b_I$.
\end{example}

\begin{example}\label{ex:2}
  Let $\mA = \mO$ be a finite field, and consider the algebraic sum
  over $\mA$. Then $a_I$ and $b_I$ are $(I,J,c_J)$-equivalent
  if $\sum\limits_{\sigma_i\in I} a_i = \sum\limits_{\sigma_i\in I} b_i$.
\end{example}

\begin{example}\label{ex:4}
  Consider the max function over the binary alphabet $\{0,1\}$. Suppose that both $I$ and $J$ are singleton, i.e., $|I|=|J|=1$. In this case, the value of $c_J$ affects the equivalence relation. When $c_J=0$, $a_I$ and $b_I$ are $(I,J,c_J)$-equivalent if $a_{I} = b_{I}$. When $c_J=1$, $a_I$ and $b_I$ are always $(I,J,c_J)$-equivalent.
\end{example}

The following properties of $(I,J,c_J)$-equivalence will be used in
our discussion.

\begin{lemma}\label{lem:red}
Let $I$ and $J$ be two disjoint subsets of $S$ and $c_J\in \mA^J$.
\begin{enumerate}
  \item When $J=\emptyset$, $(I,J,c_J)$-equivalence becomes $I$-equivalence.
  \item Let $\{J_1, J_2\}$ be a partition of $J$. For $a_I,b_I\in\mathcal{A}^{I}$ and $c_{J_1}\in\mathcal{A}^{J_1}$, if $a_I$ and $b_I$ are $(I,J_1,c_{J_1})$-equivalent, then $a_I$ and $b_I$ are $\big(I,J,(c_{J_1},c_{J_2})\big)$-equivalent for any $c_{J_2}\in\mathcal{A}^{J_2}$.
\end{enumerate}
\end{lemma}
\begin{IEEEproof}
1) is trivial. For 2), suppose $a_I$ and $b_I$ are $(I,J_1,c_{J_1})$-equivalent. Let $L_1 = S\setminus (I\cup J_1)$. We know that $f(a_I,c_{J_1},d_{L_1}) = f(b_I,c_{J_1},d_{L_1})$ for every $d_{L_1} \in \mA^{L_1}$. Let $L_2 = L_1\setminus J_2=S\setminus (I \cup J_1 \cup J_2 )$. Hence $f(a_I,c_{J_1},c_{J_2},d_{L_2}) = f(b_I,c_{J_1},c_{J_2},d_{L_2})$ for every $c_{J_2}\in \mA^{J_2}$ and $d_{L_2} \in \mA^{L_2}$, which implies that $a_I$ and $b_I$ are $(I,J,(c_{J_1},c_{J_2}))$-equivalent for any $c_{J_2}\in\mathcal{A}^{J_2}$. We complete the proof.
\end{IEEEproof}

As a special case of the above lemma, if $a_I$ and $b_I$ are $I$-equivalent, then $a_I$ and $b_I$ are $(I,J,c_J)$-equivalent for any $J\subseteq S\setminus I$ and $c_J\in\mathcal{A}^{J}$.

\subsection{Upper Bound}

Consider the network $\mN$ and the target function $f$. Let $C$ be a cut set in $\Lambda(\mN)$, and denote $I_C$ and $J_C$ by $I$ and $J$, respectively. Let $W^{(c_J)}_{I,J,f}$ be the number of the $(I,J,c_J)$-equivalence classes for a $c_J\in\mathcal{A}^{J}$ and further
$W_{C,f}=\max_{c_J \in \mathcal{A}^{J}} W_{I,J,f}^{(c_J)}$. Then, we define
\begin{IEEEeqnarray*}{rCl}
  \mincut(\mathcal{N},f) & \triangleq &
  \min_{C\in\Lambda(\mathcal{N})}\dfrac{|C|}{\log_{|\mathcal{A}|}W_{C,f}}.
\end{IEEEeqnarray*}
The following theorem shows that $\mC(\mN,f)$ is upper bounded by $\mincut(\mathcal{N},f)$, and the proof is deferred to Appendix~\ref{sec:proof-main-theorem}.

\begin{theorem}
  \label{theo1}
  For any network $\mathcal{N}$ and target function $f$,
  \[\mathcal{C}(\mathcal{N},f)\le
  \mincut(\mathcal{N},f).\]
\end{theorem}

\begin{example}
We continue to consider the network function problem $(\mN_1,f)$ in Fig.~\ref{fig:1} to illustrate this upper bound.
Let $C=\left\{ e_6,e_7 \right\}$ with $|C|=2$. Then, we have
\begin{itemize}
  \item $I_{C}=\{\sigma_3 \}$ and $J_{C}=\{\sigma_1, \sigma_2\}$, denoted by $I$ and $J$, respectively; and
  \item For any given inputs of the source nodes $\sigma_1$ and $\sigma_2$, different inputs $0$ and $1$ from the source node $\sigma_3$ output different values of $f$. Thus,
    $W_{I,J,f}^{(c_J)}=2$ for any $c_J \in \mathcal{A}^{J}$ and hence $W_{C,f}=2$.
\end{itemize}
By Theorem~\ref{theo1}, we have
$$\mathcal{C}(\mathcal{N}_1,f)\leq \mincut(\mathcal{N}_1,f)\leq\dfrac{|C|}{\log_{|\mathcal{A}|}W_{C,f}}=2.$$
On the other hand, we have shown in Section~\ref{sec:inval-prev-upper} that $ \mathcal{C}(\mathcal{N}_1,f) \geq 2$. Therefore, $\mathcal{C}(\mathcal{N}_1,f) = \mincut(\mathcal{N}_1,f) = 2$, i.e., the upper bound is tight for $(\mathcal{N}_1,f)$.
\end{example}

By 2) in Lemma~\ref{lem:red}, for any disjoint subsets $I,J\subseteq S$,
  $$R_{I,f}\geq W_{I,J,f}^{(c_J)}, \quad  \forall c_J\in \mA^{J},$$
  which implies that
  $$R_{I_C,f}\geq W_{C,f}, \quad  \forall C\in \Lambda(\mN).$$
Hence, we have in general
$$\mincut_{\text{A}}(\mathcal{N},f)\leq \mincut(\mathcal{N},f).$$

\subsection{Simplification of the Upper Bound}

In this subsection, we show that it is not necessary to explore all the cut sets in
$\Lambda(\mathcal{N})$ to determine $\mincut(\mathcal{N},f)$. As such, we can simplify the upper bound in Theorem~\ref{theo1}.

\begin{lemma}
  \label{col:red2}
  Consider a network $\mathcal{N}$ and a function $f$. Then, for any cut set $C\in
  \Lambda(\mathcal{N})$ and any subset $C'\subseteq C$ with
  $I_{C'}=I_C$, the inequality $W_{C',f}\geq W_{C,f}$ holds.
\end{lemma}
  \begin{IEEEproof}
    Let $I=I_C$, $J=J_C$ and $J'=J_{C'}$ for the notation simplicity. Apparently, $J'\subseteq J$. Let $J'' = J\setminus J'$. By Lemma~\ref{lem:red}, for any
  $c_{J'}\in\mathcal{A}^{J'}$ and $c_{J''}\in \mA^{J''}$, we have
  $$W_{I,J',f}^{(c_{J'})}\geq W_{I,J,f}^{(c_J)},$$
  where $c_J=(c_{J'}, c_{J''})$. Let $c^*_J=(c^*_{J'}, c^*_{J''})\in \mA^J$ with $c^*_{J'}\in \mA^{J'}$ and $c^*_{J''}\in \mA^{J''}$ such that $W_{I,J,f}^{(c^*_J)}=
  W_{C,f}$. Then, $W_{C',f} \geq W_{I,J',f}^{(c^*_{J'})}\geq
  W_{I,J,f}^{(c^*_J)} = W_{C,f}$. The lemma is proved.
  \end{IEEEproof}

Let $C$ be a cut set in $\Lambda(\mathcal{N})$ and $C'$ be a subset of $C$ with $I_{C'}=I_{C}$. By Lemma~\ref{col:red2}, we have $W_{C',f}\geq W_{C,f}$, which, together with $|C'|\leq |C|$, implies that
\begin{equation}\label{ineq:2}
\dfrac{|C'|}{\log_{|\mathcal{A}|}W_{C',f}}\leq\dfrac{|C|}{\log_{|\mathcal{A}|}W_{C,f}}.
\end{equation}
So it suffices to look at the ``minimum cut sets''. To be specific, for any $I\subseteq S$, define
\begin{equation*}
  \Lambda_I(\mathcal{N}) = \{C\in \Lambda(\mathcal{N}): I_C = I\},
\end{equation*}
and further
\begin{equation*}
  \Lambda_I^*(\mathcal{N}) = \{C\in \Lambda_I(\mN): C'\notin \Lambda_I(\mN),\ \forall C'\subsetneq C\},
\end{equation*}
i.e., $\Lambda_{I}^*(\mathcal{N})$ is the subset of $\Lambda_I(\mathcal{N})$ containing all $C$ in $\Lambda_{I}(\mathcal{N})$ such that no proper subsets of $C$ (i.e., subsets except itself) are in $\Lambda_I(\mathcal{N})$. Lemma~\ref{col:red2} implies
\begin{equation}\label{eq:1}
  \mincut(\mathcal{N},f) = \min_{I\subseteq S} \min_{C\in \Lambda_{I}^*(\mathcal{N})} \dfrac{|C|}{\log_{|\mathcal{A}|}W_{C,f}}.
\end{equation}

The following lemma shows that replacing
$\Lambda(\mathcal{N})$ by $\bar{\Lambda}(\mathcal{N})$ does not change
the value of $\mincut(\mathcal{N},f)$.

\begin{lemma}\label{lem:Lambda_bar}
  For any network $\mathcal{N}$ and target function $f$,
  $$
  \mincut(\mathcal{N},f) = \min_{C\in\bar{\Lambda}(\mathcal{N})}\frac{|C|}{\log_{|\mathcal{A}|}W_{C,f}}.$$
\end{lemma}
\begin{IEEEproof}
The proof of the above lemma is deferred to Appendix~\ref{sec:proof-lemma-refthm:2}.
\end{IEEEproof}

Furthermore, for any non-empty subset $I\subseteq S$, we similarly define
\begin{align}
  \bar{\Lambda}_I(\mathcal{N}) & = \{C\in \bar{\Lambda}(\mathcal{N}): I_C = I\},\label{Lambda_bar_I}\\
  \bar{\Lambda}_I^*(\mathcal{N}) & = \{C\in \bar{\Lambda}_I(\mathcal{N}): C'\notin \bar{\Lambda}_I(\mathcal{N}),\ \forall C'\subsetneq C\},\label{Lambda_bar_I_star}
\end{align}
and clearly,
\begin{equation*}
  \bar{\Lambda}_I^*(\mathcal{N}) \subseteq \Lambda_I^*(\mathcal{N}).
\end{equation*}
Now, we can give the following theorem that simplifies the upper bound in Theorem~\ref{theo1}.
\begin{theorem}\label{thm:2}
For any network $\mathcal{N}$ and target function $f$,
\begin{equation*}
  \mincut(\mathcal{N},f) = \min_{I\subseteq S} \min_{C\in \bar{\Lambda}_{I}^*(\mathcal{N})} \dfrac{|C|}{\log_{|\mathcal{A}|}W_{C,f}}.
\end{equation*}
\end{theorem}
\begin{IEEEproof}
  Using Lemma~\ref{lem:Lambda_bar} and the similar argument leading to
  \eqref{eq:1}, we obtain that
  \begin{equation*}
    \mincut(\mathcal{N},f) = \min_{C\in\bar{\Lambda}(\mathcal{N})}\frac{|C|}{\log_{|\mathcal{A}|}W_{C,f}}
    = \min_{I\subseteq S} \min_{C\in \bar{\Lambda}_{I}^*(\mathcal{N})} \dfrac{|C|}{\log_{|\mathcal{A}|}W_{C,f}},
  \end{equation*}
  which proves the theorem.
\end{IEEEproof}

\subsection{Tightness}
\label{sec:tightness-results}

We first show that our upper bound is tight for several cases and then give an example for which our bound is not tight.

\subsubsection{Single Source Node}

The network $\mathcal{N}_{\text{s}}$ has only one source node. Then, this source node can compute the function and transmits the function value to the sink node, which as a coding scheme gives
\begin{equation*}
    \mathcal{C}(\mathcal{N}_{\text{s}},f) \geq \min_{C\in \Lambda(\mathcal{N}_{\text{s}})}
    \frac{|C|}{\log_{|\mathcal{A}|}|f(\mA^s)|},
  \end{equation*}
where $f(\mA^s)$ is the set of images of $f$ and $\Lambda(\mathcal{N}_{\text{s}})$ degenerates to the set of the cuts separating this unique source node from the sink node.

Now let us check the upper bound in Theorem~\ref{theo1}. With $|S|=1$, we have $I_C=S$ and $J_C=\emptyset$, $\forall C\in \Lambda(\mathcal{N}_{\text{s}})$. Hence, $W_{C,f} = R_{I_C,f} = |f(\mA^s)|$, i.e.,
\begin{equation*}
  \mathcal{C}(\mathcal{N}_{\text{s}},f) \leq \min_{C\in \Lambda(\mathcal{N}_{\text{s}})}
    \frac{|C|}{\log_{|\mathcal{A}|}|f(\mA^s)|}.
\end{equation*}
Therefore, the upper bound is tight and
\begin{equation*}
    \mathcal{C}(\mathcal{N}_{\text{s}},f) =
    \frac{\min_{C\in \Lambda(\mathcal{N}_{\text{s}})}|C|}{\log_{|\mathcal{A}|}|f(\mA^s)|} =
    \frac{\min_{C\in \bar{\Lambda}(\mathcal{N}_{\text{s}})}|C|}{\log_{|\mathcal{A}|}|f(\mA^s)|},
  \end{equation*}
where the second equality follows from Lemma~\ref{lem:Lambda_bar}.

\subsubsection{Multi-Edge Tree}

Consider the network $\mathcal{N}_{\text{t}}$ that has a multi-edge tree topology. For two positive integers $k$ and $n$, if for each non-sink node $v$, the outgoing edges
of $v$ can transmit,  in $n$ uses, exactly $R_{I_{\eout(v)},f}^k$
messages, each of which corresponds to one
$I_{\eout(v)}$-equivalence class, i.e., $|\mA|^{n|\eout(v)|}\geq R_{I_{\eout(v)},f}^k$, then we can design a $(k,n)$ code that computes $f$ over $\mathcal{N}_{\text{t}}$ (see the proof of
\cite[Theorem~III.3]{Appuswamy11}). Therefore,
\begin{equation}\label{eq:9}
  \mathcal{C}(\mathcal{N}_{\text{t}},f) \geq \min_{v\in
    \mathcal{V}\setminus\{\rho\} } \frac{|\eout(v)|}{\log_{|\mA|}R_{I_{\eout(v)},f}}.
\end{equation}

Now let us check the upper bound in Theorem~\ref{theo1}. For any non-sink node $v$, $\eout(v)$ is a cut set in $\Lambda(\mN_{\text{t}})$ and clearly, $\eout(v)\in \bar{\Lambda}(\mN_{\text{t}})$. On the other hand, for any edge-subset $E$ satisfying that $\eout(v) \nsubseteq E$, $\forall v\in\mV\setminus \{ \rho \}$, we see that
$I_E=\emptyset$ and hence $E$ is not a cut-set.
Therefore,
\begin{equation}\label{eq:2}
   \{\eout(v): v \in \mathcal{V}\setminus\{\rho\} \} = \bigcup_{I\subseteq S}\bar{\Lambda}_{I}^*(\mathcal{N}_{\text{t}}).
\end{equation}
By Theorems~\ref{theo1} and \ref{thm:2}, we have
\begin{equation}\label{eq:10}
  \mathcal{C}(\mathcal{N}_{\text{t}},f) \leq \min_{I\subseteq S} \min_{C\in
    \bar{\Lambda}_{I}^*(\mathcal{N}_{\text{t}})}
  \dfrac{|C|}{\log_{|\mathcal{A}|}W_{C,f}} = \min_{v\in
    \mathcal{V}\setminus\{\rho\} } \frac{|\eout(v)|}{\log_{|\mA|}W_{{\eout(v)},f}},
\end{equation}
where the equality follows from \eqref{eq:2}. In addition, since the network $\mathcal{N}_{\text{t}}$ has a multi-edge tree topology, we have $J_{\eout(v)} =\emptyset$, $\forall v \in \mathcal{V}\setminus\{\rho\}$. This implies $W_{{\eout(v)},f} = R_{I_{\eout(v)},f}$ (see Lemma~\ref{lem:red}-1)). Therefore, combining \eqref{eq:9} with \eqref{eq:10}, the
upper bound in Theorem~\ref{theo1} is tight for
$\mathcal{N}_{\text{t}}$, i.e.,
\begin{equation*}
  \mathcal{C}(\mathcal{N}_{\text{t}},f) = \min_{v\in
    \mathcal{V}\setminus\{\rho\} } \frac{|\eout(v)|}{\log_{|\mA|}R_{I_{\eout(v)},f}}.
\end{equation*}

\subsubsection{Identity Function}

Let $\mathcal{A}^s=\mathcal{O}$ and define the
identify function $f_{\mathrm{id}}:\mathcal{A}^s\rightarrow
\mathcal{O}$ as $f_{\mathrm{id}}(\bx)=x$ for all $x\in \mathcal{A}^S$.  When the target
function is the identity function $f_{\mathrm{id}}$, the network
function computation problem becomes the multiple access problem, where the sink
node is required to recover the symbols generated by all the source nodes.
From \cite[Theorem~4.2]{lehman}, we know that \begin{equation*}
  \mathcal{C}(\mathcal{N},f_{\mathrm{id}}) = \min_{C\in
    \Lambda(\mathcal{N})} \frac{|C|}{|I_C|},
\end{equation*}
and \emph{forwarding}\footnote{We say a network code is \emph{forwarding} if for each intermediate
node $u \in \mathcal{V} \setminus (S\cup\{\rho\})$ and each edge $e\in
\eout(u)$, the symbols transmitted on edge $e$ are all in the set of symbols received by node $u$ from its incoming edges.} is sufficient to achieve the capacity.
For the identity function $f_{\mathrm{id}}$, $a_I$ and $b_I$ in $\mA^I$ are $(I,J,c_J)$-equivalent if and only if $a_I=b_I$ no matter what $c_J$ is in $\mA^J$. Therefore, we have $W_{C,f} = |\mA|^{|I_C|}$, $\forall C\in \Lambda(\mN)$, which gives
\begin{equation*}
  \mincut(\mathcal{N},f_{\mathrm{id}}) = \min_{C\in\Lambda(\mathcal{N})} \frac{|C|}{|I_C|} = \mathcal{C}(\mathcal{N},f_{\mathrm{id}}).
\end{equation*}

\subsubsection{Algebraic Sum over a Finite Field}
\label{sec:linear-functions}

Suppose $\mathcal{A}=\mathcal{O}$ is a finite field. The algebraic sum
function $f_{\text{sum}}:\mathcal{A}^s\rightarrow \mathcal{A}$ is defined as
$f_{\text{sum}}(x_1,x_2,\cdots,x_s) = \sum_{i=1}^s x_i$. We know from \cite{koetter2004network} that
\begin{equation*}
  \mathcal{C}(\mathcal{N},f_{\text{sum}}) = \min_{\sigma\in S} \min_{C\in
    \Lambda(\mathcal{N}): \sigma\in I_C } |C|=\min_{C\in
    \Lambda(\mathcal{N})} |C|,
\end{equation*}
and linear network coding is sufficient to achieve this capacity, and particularly, such a linear network code can be obtained by using the duality relation between the sum networks and the multicast networks.

For the algebraic sum $f_{\text{sum}}$, the number of the $(I,J,c_J)$-equivalence classes is always $|\mA|$ for any disjoint subsets $I$ and $J$ of $S$ and any $c_J\in \mA^J$. Hence, $W_{C,f} = |\mA|$. Therefore,
\begin{equation*}
  \mincut(\mathcal{N},f_{\mathrm{sum}}) = \min_{C\in\Lambda(\mathcal{N})} |C| = \mathcal{C}(\mathcal{N},f_{\text{sum}}).
\end{equation*}

\subsubsection{Looseness}

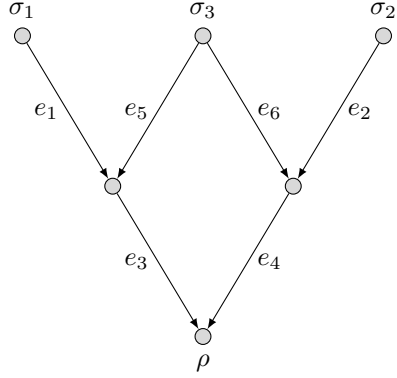
\begin{figure}
  \centering
 \begin{tikzpicture}[x=0.6cm]
    \draw (0,0) node[vertex] (3) [label=above:$\sigma_3$] {};
    \draw (-2,-2) node[vertex] (1) {};
    \draw (2,-2) node[vertex] (2) {};
    \draw (0,-4) node[vertex] (0) [label=below:$\rho$] {};
    \draw[->,>=latex] (3) -- (1) node[midway, left] {$e_5$};
    \draw[->,>=latex] (3) -- (2) node[midway, right] {$e_6$};
    \draw[->,>=latex] (1) -- (0) node[midway, left] {$e_3$};
    \draw[->,>=latex] (2) -- (0) node[midway, right] {$e_4$};

    \node[vertex,label=above:$\sigma_1$] at (-4,0) {} edge[->,>=latex] node[midway,left] {$e_1$} (1);
    \node[vertex,label=above:$\sigma_2$] at (4,0) {} edge[->,>=latex] node[midway,right] {$e_2$} (2);
\end{tikzpicture}

\caption{Network $\mathcal{N}_2$ has three
  binary sources $\sigma_1$, $\sigma_2$, $\sigma_3$ and one sink node $\rho$. The arithmetic sum of the source messages is required to be computed, where
  $\mathcal{A}=\mathcal{B}=\left\{0,1 \right\}$ and $\mathcal{O}=\left\{0,1,2,3 \right\}$.}
  \label{fig:5}
\end{figure}

The upper bound in Theorem~\ref{theo1} is not tight in general. Consider the network $\mathcal{N}_2$ in
Fig.~\ref{fig:5}, where the arithmetic sum $f_{\text{a-sum}}$ is required to be calculated at the sink node. It has been proved in \cite{Appuswamy11} that
$\mathcal{C}(\mathcal{N}_2,f_{\text{a-sum}})=\log_64$. But Theorem~\ref{theo1} shows by a simple calculation that $$\mincut(\mathcal{N}_2,f_{\text{a-sum}}) = 1> \log_64 =\mathcal{C}(\mathcal{N}_2,f_{\text{a-sum}}).$$

Following the conference version of this paper, an improved upper bound is proposed
recently in \cite{xuan16computing} by applying a refined equivalence relation and a cut-set
partition, which is tight for computing arithmetic sum over $\mathcal{N}_2$.

\section{Further Comments}
\label{sec:further-comments}

Some results in \cite{Appuswamy13, Appuswamy14} were proved based on
the invalid upper bound $\mincut_{\text{A}}(\mathcal{N},f)$ in
\cite{Appuswamy11}. In particular, $\mincut_{\text{A}}(\mathcal{N},f)$
is used as an upper bound on the computing capacity in the proofs of
Theorem IV.5, Lemma V.6 and Theorem V.2 in \cite{Appuswamy13}, and
Lemma II.5 in \cite{Appuswamy14}. Since
$\mincut_{\text{A}}(\mathcal{N},f)$ is not a valid upper bound in
general, the correctness of these results needs further justification.
In this section, we verify that these results are all correct.

As we have mentioned before, $(I,J,c_J)$-equivalence becomes
$I$-equivalence in some cases so that $\mincut(\mathcal{N},f) =
\mincut_{\text{A}}(\mathcal{N},f)$.
In Examples~\ref{ex:1} and~\ref{ex:2}, we see that $(I,J,c_J)$-equivalence can be the same as
$I$-equivalence for the identity function and the algebraic sum function.\footnote{Instead, Example~\ref{ex:4} shows that
$(I,J,c_J)$-equivalence can be different from $I$-equivalence.}
Therein, the validity of \cite[Theorem IV.5 and Lemma V.6]{Appuswamy13} is implied by Example~\ref{ex:2} and Theorem~\ref{theo1}. In addition, the following examples show that for arithmetic sum functions and vector linear functions, $(I,J,c_J)$-equivalence is also the same as $I$-equivalence.

\begin{example}\label{ex:3}
  Suppose $\mA=\{0,1\}$ and $\mO=\{0,1,2,\cdots,s\}$, and consider arithmetic sum function $f:\mathcal{A}^s\rightarrow \mathcal{O}$. Then $a_I$ and $b_I$ in $\mA^I$ are $(I,J,c_J)$-equivalent
  if $\sum_{\sigma_i\in I} a_{i} = \sum_{\sigma_i\in I} b_{i}$. The current example, together with Theorem~\ref{theo1}, implies the validity of \cite[Theorem V.2]{Appuswamy13}.
\end{example}

\begin{example}
\label{lem:linear}
Let $\mA$ be a finite field $\Fq$ and $\mO =
\Fq^l$, where $q$ is a prime power. Consider a
  linear function $f: \Fq^s\to \Fq^l$ defined by $f(x)=x\cdot T$, where $T$ is an $s\times l$ matrix over $\Fq$.
Then for disjoint $I,J\subseteq S$, and $c_J\in\mathcal{A}^{J}$,
$(I,J,c_J)$-equivalence is equivalent to $I$-equivalence, which together with Theorem~\ref{theo1} implies the validity of \cite[Lemma II.5]{Appuswamy14}.
\end{example}

\appendices

\section{Proof of Theorem~\ref{theo1}}
\label{sec:proof-main-theorem}

We first define some notations and prove a lemma.
Given a network $\mathcal{N}$ and a cut set
$C\in\Lambda(\mathcal{N})$, define 
  $$D(C)=\bigcup_{ \sigma\in S\setminus I_C}\eout(\sigma),$$
and let
  $$F(C)=C\cup D(C).$$
It follows from $C\subseteq F(C)$ that $I_C\subseteq I_{F(C)}$. On the other hand, since $K_{\eout(\sigma)} = I_{\eout(\sigma)}=\{\sigma\}$, we have $I_{D(C)}=S\setminus I_C$. Therefore,
\begin{align*}
S=I_C\cup I_{D(C)}\subseteq I_{C\cup D(C)}=I_{F(C)}\subseteq S,
\end{align*}
which shows that for every cut set $C$, $F(C)$ is a global cut set, i.e.,
$$I_{F(C)}=S,\quad \forall C\in\Lambda(\mathcal{N}).$$

\begin{lemma}
  \label{lem2}
  Consider a $(k,n)$ network code for computing $f$ over $\mathcal{N}=(G,S,\rho)$. For any global cut set $C$, the decoding function
  $\psi(x_S)$ of this $(k,n)$ network code is a function of $g_{C}(x_S)$, i.e., for all $x_S, y_S\in \mA^{k\times S}$, $\psi(x_S) =\psi(y_S)$ provided that
 $g_C(x_S)=g_C(y_S)$.
\end{lemma}
\begin{IEEEproof}
Recall that $\psi(x_S)$ is a function of $g_{\ein(\rho)}(x_S)$. It
suffices to prove that $\forall x_S, y_S\in \mA^{k\times S}$,
$g_{\ein(\rho)}(x_S) = g_{\ein(\rho)}(y_S)$ provided that
$g_C(x_S)=g_C(y_S)$. This will be proved by contradiction. Suppose
$\exists a_S, b_S\in \mA^{k\times S}$ such that
\begin{align}\label{equ:1_in_App1}
g_{\ein(\rho)}(a_S) \neq g_{\ein(\rho)}(b_S),
\end{align}
but
\begin{align}\label{equ:2_in_App1}
g_C(a_S)=g_C(b_S).
\end{align}
By \eqref{equ:1_in_App1}, $\exists e\in \ein(\rho)$ such that
\begin{align}\label{equ:3_in_App1}
g_{e}(a_S) \neq g_{e}(b_S).
\end{align}
We claim $e\notin C$ because otherwise \eqref{equ:3_in_App1} contradicts to \eqref{equ:2_in_App1}.

Furthermore, it follows from \eqref{equ:3_in_App1} that $g_{\ein(\tail(e))}(a_S) \neq g_{\ein(\tail(e))}(b_S)$ since $g_e(x_S)=h_e\big(g_{\ein(\tail(e))}(x_S)\big)$, where $h_e$ is the local encoding function for $e$. By the same argument, $\exists d\in \ein(\tail(e))$ such that $g_{d}(a_S) \neq g_{d}(b_S)$ and $d\notin C$. So on and so forth, since the cut set $C$ is global and the graph $G$ is finite, there must exist $e'\in C$ such that $g_{e'}(a_S) \neq g_{e'}(b_S)$, contradicting to \eqref{equ:2_in_App1}. Then we accomplish the proof.
\end{IEEEproof}

To prove Theorem~\ref{theo1}, it will be handy to generalize the equivalence
relation for a block of function inputs. Let $k$ be a positive integer. For two disjoint sets $I,J\subseteq S$ and $c\in\mathcal{A}^{J}$ we say $a,b\in\mathcal{A}^{k\times
  I}$ are $(I,J,c)$-equivalent if for any
$x_S,y_S\in\mathcal{A}^{k\times S}$, respectively written as $x_S=(x_I, x_J, x_{S\setminus (I\cup J)})$ and $y_S=(y_I, y_J, y_{S\setminus (I\cup J)})$, with $x_I=a,y_I=b,x_J=y_J=\left(c^\top,c^\top,\dots,c^\top \right)^{\top}$ and  $\bx_{S\backslash I\cup J}=y_{S\backslash I\cup
  J}$, it is always that $f(x_S)=f(y_S)$. Then for the set
$\mathcal{A}^{k\times I}$, the number of the equivalence classes induced by this equivalence relation is $\left( W_{I,J,f}^{(c)}\right)^k$.

\begin{IEEEproof}[Proof of Theorem~\ref{theo1}]
Suppose we have a $(k,n)$ code with
\begin{equation}
  \label{eq:5}
  \frac{k}{n} \log_{|\mathcal{B}|}|\mathcal{A}|>
\mincut(\mathcal{N},f).
\end{equation}
We show that this code cannot compute $f(\bx)$ correctly for all $x\in
\mathcal{A}^{k\times S}$. Let $C^*$ be a cut set in $\Lambda(\mathcal{N})$ such that
\begin{equation}
  \label{eq:6}
  \dfrac{|C^*|}{\log_{|\mathcal{A}|}W_{C^*,f}}
  =\min_{C\in\Lambda(\mathcal{N})}\dfrac{|C|}{\log_{|\mathcal{A}|}W_{C,f}}.
\end{equation}
  Further, let $I^*=I_{C^*}$, $J^*=J_{C^*}$ and $c^*\in \mathcal{A}^{J^*}$ such that
  \begin{equation}
    \label{eq:7}
    W^{(c^*)}_{I^*,J^*,f}=\max_{c\in\mathcal{A}^{J^*}}W^{(c)}_{I^*,J^*,f}.
  \end{equation}
  By \eqref{eq:5}-\eqref{eq:7}, we have
  $$\frac{k}{n}\log_{|\mathcal{B}|}|\mathcal{A}|>
  \frac{|C^*|}{\log_{|\mathcal{A}|}W_{I^*,J^*,f}^{(c^*)}},$$
  which leads to
  \begin{equation}\label{eq:3}
    |\mathcal{B}|^{n|C^*|}<\left(W^{(c^*)}_{C^*,f}\right)^{k}.
  \end{equation}

Note that $g_{C^*}(x_S)$ only depends on $(x_{I^*}, x_{J^*})$, the inputs generated by the source nodes in $I^*\cup J^*$. By \eqref{eq:3} and the pigeonhole principle, there exist $a, b \in \mathcal{A}^{k\times I^*}$ such that the following are qualified:
\begin{enumerate}
  \item $a$ and $b$ are not $(I^*,J^*,c^*)$-equivalent; and
  \item $g_{C^*}(x_S)=g_{C^*}(y_S)$ for
any $x_S=(x_{I^*}, x_{J^*}, x_{S\setminus (I^*\cup J^*)}),y_S=(y_{I^*}, y_{J^*}, y_{S\setminus (I^*\cup J^*)})\in\mathcal{A}^{k\times S}$ with
\begin{equation}
  \label{eq:4} \left\{
  \begin{array}[]{l}
    x_{I^*} =a,\quad y_{I^*}= b, \\
    x_{J^*} = y_{J^*}=(c^{*\top},c^{*\top},\dots,c^{*\top})^\top\triangleq c\in \mA^{k\times J^*}, \\
    x_{S\setminus (I^*\cup J^*)} = y_{S\setminus (I^*\cup J^*)}.
  \end{array}
  \right.
\end{equation}
\end{enumerate}

Thus, by 1), $\exists d\in \mA^{k\times S\setminus (I^*\cup J^*)}$ such that $f(a,c,d)\neq f(b,c,d)$. Let $x_S=(a,c,d)$ and $y_S=(b,c,d)$. Then by 2), we have $g_{C^*}(x_S)=g_{C^*}(a, c)=g_{C^*}(b, c)=g_C(y_S)$. Note that $C^*\cap D(C^*)=\emptyset$. We obtain
\begin{align*}
  g_{F(C^*)}(x_S)=&\big( g_{C^*}(a, c), g_{D(C^*)}(c, d) \big)= \big( g_{C^*}(b, c), g_{D(C^*)}(c, d) \big)= g_{F(C^*)}(y_S).
\end{align*}
Then, it follows from Lemma~\ref{lem2} that $\psi(x_S)=\psi(y_S)$. Therefore, the code cannot correctly computes either $f(x_S)$ or $f(y_S)$. The proof is completed.
\end{IEEEproof}

\section{Proof of Lemma~\ref{lem:Lambda_bar}}
\label{sec:proof-lemma-refthm:2}

To prove Lemma~\ref{lem:Lambda_bar}, we first generalize some notations about cut sets and prove a graph-theoretic lemma.

For a finite DAG $G=(\mathcal{V},\mathcal{E})$, let $S$ and $T$ be two
disjoint subsets of $\mathcal{V}$. An edge-subset $C\subseteq
\mathcal{E}$ is called an \emph{$S$-$T$ cut set} if any
node in $S$ is separated from any node in $T$ upon deleting the edges
in $C$ from $\mathcal{E}$. An $S$-$T$ cut set $C$ is said to be minimum, if $C$ has the minimum cardinality among all $S$-$T$ cut sets. A node-subset $U\subseteq \mathcal{V}$ is
said to be an \emph{$S$-$T$ cut} if $S\subseteq U$ and $T\subseteq
\mV\setminus U$. For an $S$-$T$ cut $U$, if $\mE(U)$ is a minimum $S$-$T$ cut set, $U$ is called a {\em minimum $S$-$T$ cut}.

For $u,v \in U \subseteq \mathcal{V}$, if there exists
a directed path from $u$ to $v$ and all nodes on this path are in $U$, we say $v$ is reachable in $U$ by $u$. We adopt the convention that any node $u\in U$ is reachable in $U$ by itself.
For an edge-subset $E\subseteq \mE$, we define two node-subsets
$$\tail(E)=\{ \tail(e): e\in E \} \quad \text{and} \quad \head(E)=\{ \head(e): e\in E \},$$
which are the sets of tail and head nodes of all edges in $E$, respectively.

\begin{lemma}\label{lem:app}
Let $G=(\mathcal{V},\mathcal{E})$ be a finite DAG, and $S$ and $T$ be
two disjoint subsets of $\mathcal{V}$. Then there exists a minimum $S$-$T$ cut $U$ such that each node in $U$ is reachable in $U$ by a node in $S$.
\end{lemma}
\begin{IEEEproof}
Let $V\subseteq \mV$ be a minimum $S$-$T$ cut. Let $U\subseteq \mV$ be
the subset of nodes each of which is reachable by at least one node in
$S$ upon deleting the edges in $\mE(V)$. By the construction of $U$,
we have
\begin{enumerate}
\item each node in $U$ is reachable in $U$ by a node in $S$;
\item $S\subseteq U$; and
\item $T\subseteq \mathcal{V}\setminus U$, since otherwise there exists
  a path from a node in $S$ to a node in $T$ that does not use any
  edge in $\mE(V)$, a contraction to the fact that $V$ is an $S$-$T$ cut.
\end{enumerate}
In the next paragraph, we will prove  $\mE(U)\subseteq \mE(V)$
which, together with 2) and 3), implies that $U$ is also a minimum
$S$-$T$ cut, proving the lemma.

Assume the contrary that
there is an edge $d\in \mE(U)$ but $d\notin \mE(V)$. Note that $d\in \mE(U)$ implies $\tail(d)\in U$. Thus, by the construction of $U$,
there exists a directed path $P$ from a node $s$ in $S$ to $\tail(d)$ and $P$ does not go through any edge in $\mE(V)$.
Moreover, since $d\notin \mE(V)$, we can extend $P$ to node
$\head(d)$ by appending $d$, so that $\head(d)$ is reachable by $s$ upon deleting the edges in $\mE(V)$. In other words, $\head(d)\in U$ by the construction of $U$. This is a contradiction to $d\in \mE(U)$.  The
lemma is proved.
\end{IEEEproof}

Now, we prove Lemma~\ref{lem:Lambda_bar}.

\begin{IEEEproof}[Proof of Lemma~\ref{lem:Lambda_bar}]
 It follows immediately from $\bar{\Lambda}(\mathcal{N})\subseteq \Lambda(\mathcal{N})$ that
$$\mincut(\mathcal{N},f)\leq \min_{C\in\bar{\Lambda}(\mathcal{N})}\frac{|C|}{\log_{|\mathcal{A}|}W_{C,f}}.$$
So it suffices to prove another direction of the inequality.

Consider a cut set $C$ in $\Lambda(\mathcal{N})$ such that
\begin{align}\label{thm2:pf_eq7}
\dfrac{|C|}{\log_{|\mathcal{A}|}W_{C,f}}=\mincut(\mathcal{N}, f).
\end{align}
By Lemma~\ref{lem:app}, we can let $U\subseteq \mV$ be a minimum $I_C$-$\head(C)$ cut such that each node in $U$ is reachable in $U$ by a source node in $I_C$. Then, we have
\begin{align}
|\mE(U)| & \leq  |C|,\label{thm2:pf_eq5} \\
I_{\mE(U)} & \supseteq I_C. \label{eq:ddsl}
\end{align}

We claim that for each edge $d\in \mE(U)$, either $d\in C$ or every directed path from $\head(d)$ to $\rho$ passes through at least one edge in $C$. To see this,
we assume the contrary that there exists an edge $d\in \mE(U)\setminus C$ (if non-empty) with a directed path from $\head(d)$ to $\rho$ not passing through any edge in $C$. This path is denoted by $P^{\head(d)\rightarrow \rho}$. On the other hand, since $\tail(d)$ is reachable in $U$ by a source node in $I_C$, say $\sigma$, there is a directed path from $\sigma$ to $\tail(d)$ such that all nodes on the path are in $U$. Denote this path by $P^{\sigma \rightarrow \tail(d)}$. Further, note that no edges in $C$ are on $P^{\sigma \rightarrow \tail(d)}$, because otherwise $\head(C)\cap U\neq \emptyset$. We concatenate $P^{\sigma \rightarrow \tail(d)}$, $d$ and $P^{\head(d)\rightarrow \rho}$ in order and then obtain a path from $\sigma$ to $\rho$ not passing through any edge in $C$, a contradiction to $\sigma\in I_C$. 

The above claim implies that $J_{\mE(U)}\subseteq J_C$ and $I_{\mE(U)}\subseteq I_C$, which together with \eqref{eq:ddsl} gives $I_{\mE(U)}=I_C$.
By Lemma~\ref{lem:red}, we further have
\begin{equation*}
W_{\mE(U),f}\geq W_{C,f},
\end{equation*}
which together with \eqref{thm2:pf_eq5} gives
\begin{equation}
\label{thm2:pf_eq6}
\dfrac{|\mE(U)|}{\log_{|\mathcal{A}|}W_{\mE(U),f}}\leq \dfrac{|C|}{\log_{|\mathcal{A}|}W_{C,f}}.
\end{equation}
Then we obtain
\begin{equation*}
\min_{C\in\bar{\Lambda}(\mathcal{N})}\dfrac{|C|}{\log_{|\mathcal{A}|}W_{C,f}}
\leq \dfrac{|\mE(U)|}{\log_{|\mathcal{A}|}W_{\mE(U),f}}
\leq \dfrac{|C|}{\log_{|\mathcal{A}|}W_{C,f}}=\mincut(\mathcal{N},f),
\end{equation*}
where the second inequality and the last equality follow from
\eqref{thm2:pf_eq6} and \eqref{thm2:pf_eq7}, respectively.
\end{IEEEproof}

\section*{Acknowledgement}
This work was partially supported by NSFC Grant (Nos.
61471215, 61301137), the University Grants Committee of
the Hong Kong SAR, China (Project No. AoE/E-02/08), and
the Vice-Chancellor's One-off Discretionary Fund of CUHK
(Project Nos. VCF2014030 and VCF2015007).

\bibliographystyle{IEEEtran}
\end{document}